\documentclass{egpubl}
\usepackage{STAG2022P}
 
 \Poster                 
\WsPoster          

\usepackage[T1]{fontenc}
\usepackage{dfadobe}  
\usepackage{float} 
\usepackage{amsmath}
\usepackage{amssymb}

\usepackage{cite}  
\BibtexOrBiblatex
\electronicVersion
\PrintedOrElectronic
\ifpdf \usepackage[pdftex]{graphicx} \pdfcompresslevel=9
\else \usepackage[dvips]{graphicx} \fi

\usepackage{egweblnk} 

\title[Floor Plan Exploration Framework Based on Similarity Distances]%
      {Floor Plan Exploration Framework Based on Similarity Distances}

\author[Chia-Ying Shih \& Chi-Han Peng]
{\parbox{\textwidth}{\centering Chia-Ying Shih$^1$ and Chi-Han Peng$^1$\orcid{0000-0002-6823-8029}}
        \\
{\parbox{\textwidth}{\centering National Yang Ming Chiao Tung University$^1$}
}
}

\begin{document}

\maketitle
\begin{abstract}
Computational methods to compute similarities between floor plans can help architects explore floor plans in large datasets to avoid duplication of designs and to search for existing plans that satisfy their needs. Recently, LayoutGMN~\cite{patil2021layoutgmn} delivered state-of-the-art performance for computing similarity scores between floor plans. However, the high computational costs of LayoutGMN make it unsuitable for the aforementioned applications. In this paper, we significantly reduced the times needed to query results computed by LayoutGMN by projecting the floor plans into a common low-dimensional (e.g., three) data space. The projection is done by optimizing for coordinates of floor plans with Euclidean distances mimicking their similarity scores originally calculated by LayoutGMN. Quantitative and qualitative evaluations show that our results match the distributions of the original LayoutGMN similarity scores. User study shows that our similarity results largely match human expectations.

\end{abstract}  
\section{Introduction}

To search for similar floor plan designs to copy or to avoid, architectural designers often spend a lot of time exploring floor plans in massive datasets such as RPLAN (\cite{wu2019data}). However, most floor plan similarity calculation methods (including traditional methods such as Intersection-over-Union (IoU) and neural network-based methods such as LayoutGMN~\cite{patil2021layoutgmn}) are not fast enough to support searching in interactive speeds. Therefore, we propose an efficient approach for the goal. In short, we project floor plans into a common low-dimensionality data space such that their Euclidean distances would mimic the similarity scores computed by LayoutGMN, a state-of-the-art floor plan similarity calculation method. In this data space, data exploration tasks such as: 1) searching for similar floor plans, 2) clustering, and 3) pruning of nearly redundant designs, can be efficiently conducted. For example, our method reduced the computation time for task 1) from 296.8 seconds by the original LayoutGMN to just 8.7 seconds. To verify if the similarity measures computed by our method match human expectations, we conducted user studies in questionnaire formats. The results showed that the respondents largely agree with our method's similarity estimations.
    
    %
    


\noindent \textbf{Related Work.} Determining the similarity between floor plans is a non-trivial task because not only the shapes but also the connectivity of rooms need to be considered. Such "distance" measures between floor plans is a key component for generative methods~\cite{nauata2020house}. Early approaches only search for specific areas (such as bedroom and living room)~\cite{kiyota2005simulated}. Recent methods tend to learn the overall spatial features ~\cite{sharma2017daniel} and to present floor plans as planar graphs encoding the rooms as vertices~\cite{patil2021layoutgmn}. Recently, LayoutGMN~\cite{patil2021layoutgmn} produced state-of-the-art quantitative and qualitative performances of similarity computation between floor plans by encoding room characteristics (e.g., aspect ratios and locations) and room relationship (e.g., the relative positions and overlapping areas) using an attension-based graph matching network. However, LayoutGMN only computed tripled similarities in which each floor plan don't have a unique feature vector. This means that we cannot simply project the floor plans to a common data space using the feature vectors as coordinates. Another shortcoming of LayoutGMN is that the similarity scores are not upper-bounded, further complicating the data projection task.
\begin{figure}[H]
\centering
\includegraphics[width=1\linewidth]{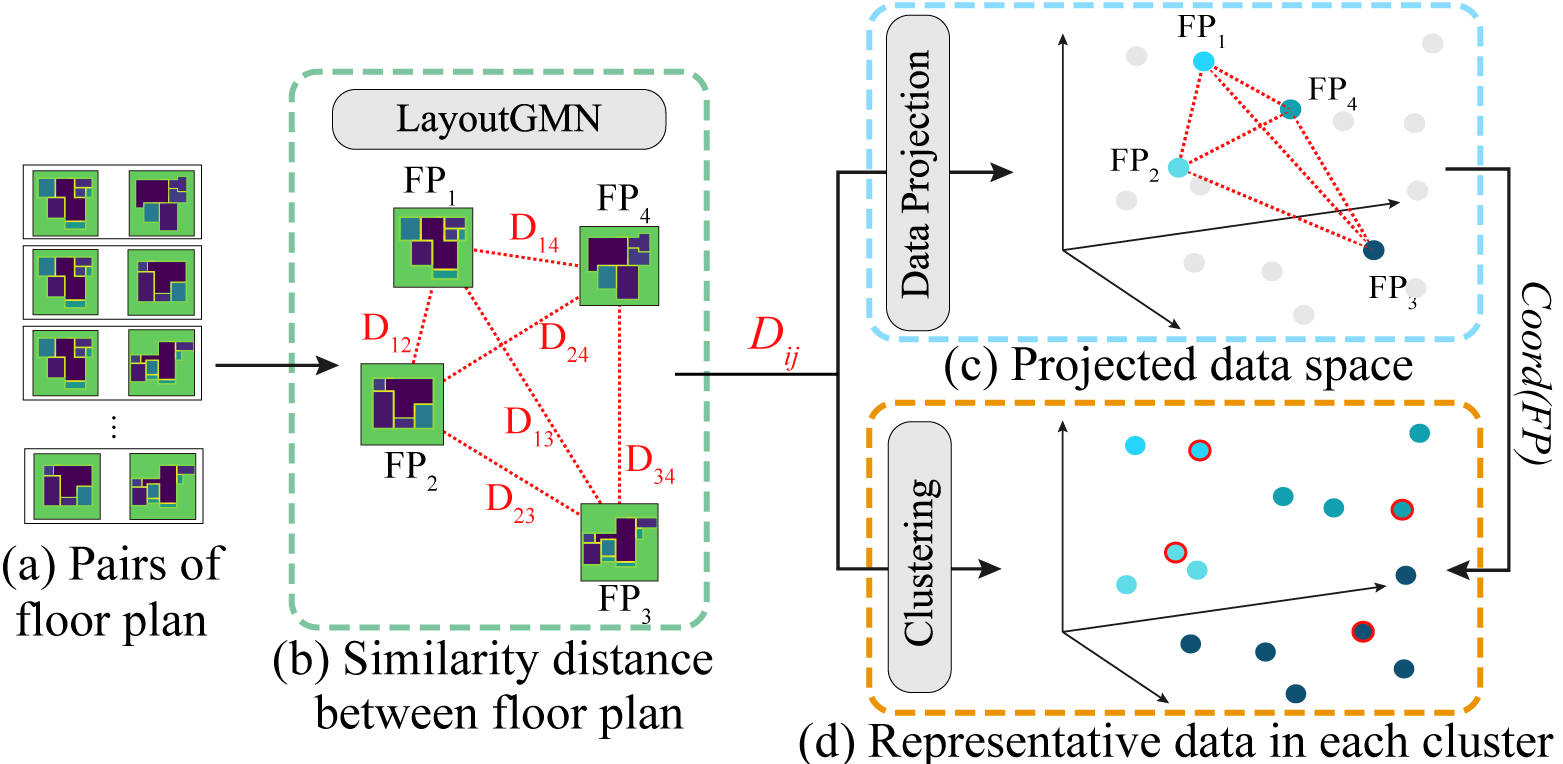}
\caption{Overview of our method. (b) Original pair-wise similarity scores between floor plans are computed by LayoutGMN. (c) The floor plans are projected to a common low-dimensional space such that the Euclidean distances mimic the similarity scores, making data exploration tasks such as clustering (d) easier to perform.}
\label{fig:our_architecture}
\end{figure}

     
     
    
\section{Method}

Our method is summarized in Figure ~\ref{fig:our_architecture}. We denote floor plans as $FP_i$, $ 0 \le i < N$, $N$ is the number of floor plans in the dataset. Recall that LayoutGMN compute similarity between floor plans by computing features vectors (1024 in length) of triples of floor plans and then computing Euclidean distances of the feature vectors as the similarity scores. Note that the feature vector of a floor plan could be different when it is included in different triples. Therefore, they could not be directly used as coordinates in a common 1024-dimensional space. Through experiments, we have found that directly applying traditional dimensionality-reduction techniques such as PCA produced bad results because LayoutGMN tends to produce similarity scores of very large value ranges. Therefore, we instead use cosine similarity to encode distances between two floor plans into the 0 to 1 range as follows:
\begin{equation}
Dist(FP_i,FP_j) = \frac{ V_{i,(i,j,k)} \cdot V_{j,(i,j,k)} }{ \lVert V_{i,(i,j,k)} \rVert * \lVert V_{j,(i,j,k)} \rVert } * -1 + 1
\end{equation}
where $V_{i,(i,j,k)} \in R^{1024} $ is the feature vector of floor plan $FP_i$ when calculated among a triple of floor plans $FP_i$, $FP_j$, and $FP_k$ by LayoutGMN. Note that LayoutGMN selects the triples such that each triple would include one "anchor" floor plan, one "positive" floor plan that has a higher IoU score to the anchor, and one "negative" one that has a lower IoU score to the anchor, respectively. The similarity score of two floor plans is guaranteed to be the same even if the pair is included in different triples because the pair is still processed by the same GMN module.

Next, we solve for the coordinates of the floor plans in a common data space as a linear least squares optimization problem formulated as follows:
\begin{equation}
\label{equ:optimization}
\begin{aligned}
& \underset{X_i, \ 0 \le i < N}{\text{argmin}}
& & \sum_{\forall Dist(FP_i,FP_j) \, exists}{(\lVert (X_i - X_j) \rVert - Dist(FP_i,FP_j))^2}
\end{aligned}
\end{equation}
where $X_i$ is the coordinate of $FP_i$.

    
\section{Results and Applications}

\begin{figure}[]
        \centering
        \includegraphics[width=1\linewidth]{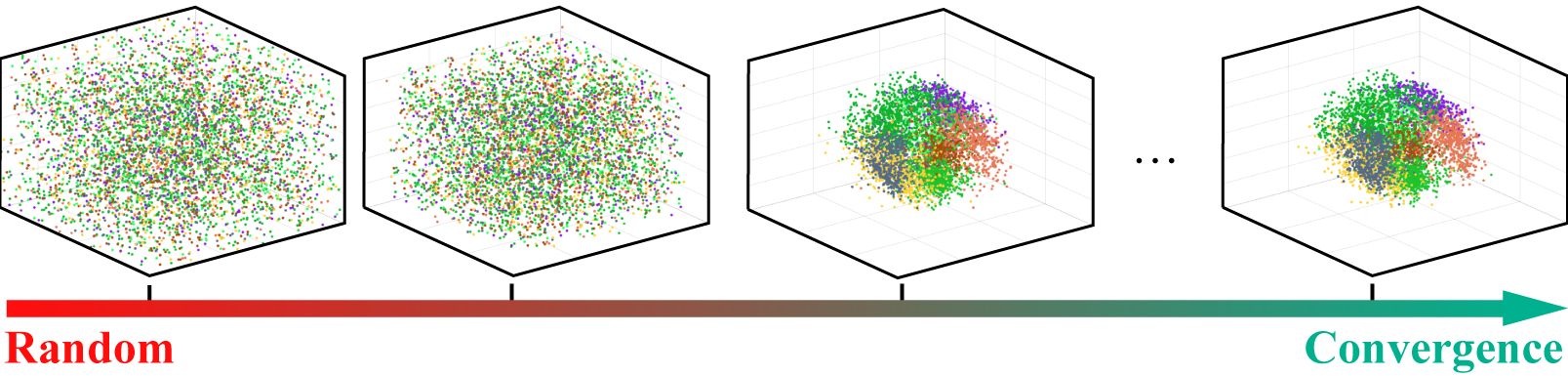}
        \caption{Projected 3D coordinates of floor plans and their clustering according to the original LayoutGMN similarity scores.}
        \label{fig:iteration_projection}
    \end{figure}

We tested on a computer with Intel i9-10900 2.80GHz CPU and 32GB Rams. We trained LayoutGMN on the whole RPLAN dataset (77664 floor plans) to get 12492731 pairs of similarity scores, and used Google Ceres solver to solve the projected coordinates of the floor plans. We can solve up to 3D data spaces due to memory constraints. Quantitatively speaking, the solving using cosine similarity (which took 56.65 minutes to converge) reduced the residuals from 1.9e+06 (random) to 1.3e+03, hinting a successful convergence. In comparisons, the solving could not converge successfully using original Euclidean distances or log distances. To qualitatively evaluate the projection results, in Figure ~\ref{fig:iteration_projection}, we show iterations of intermediate solutions of projected coordinates. By coloring the data points with clustering of the similarity scores calculated by LayoutGMN, we find that our projected coordinates largely match the distributions of the original similarity scores.

\begin{figure}[]
        \centering
        \includegraphics[width=1\linewidth]{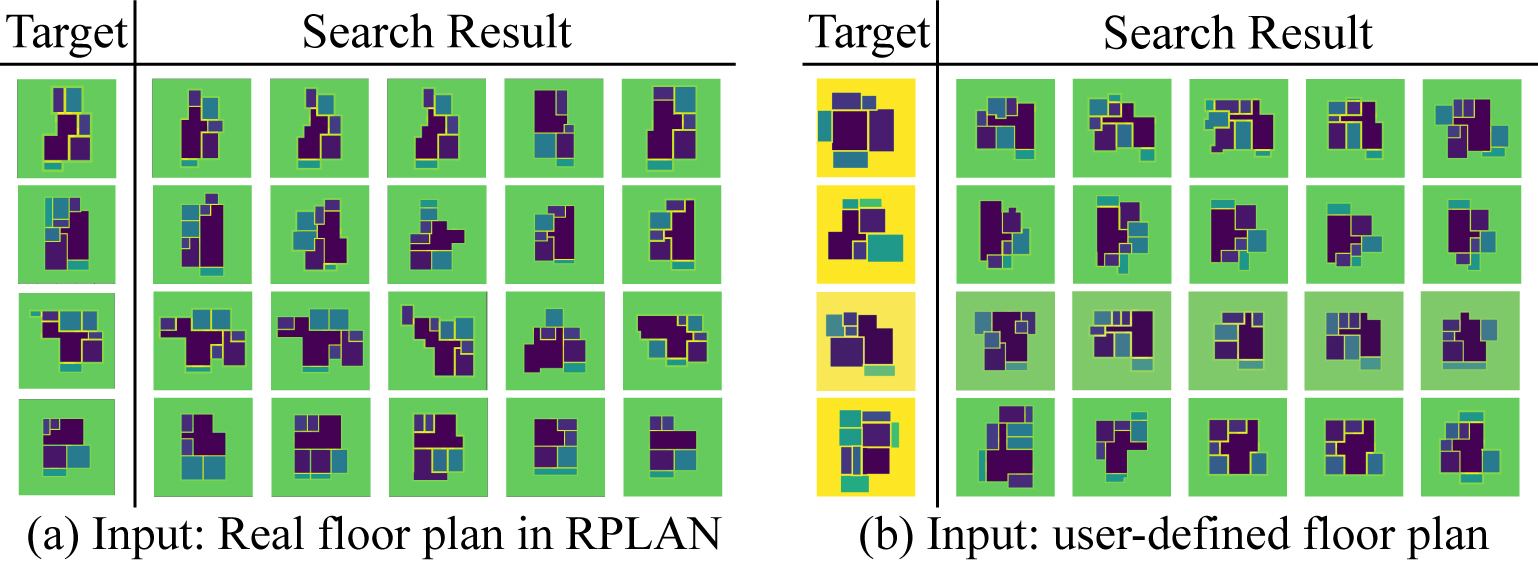}
        \caption{Searching similar designs by existing (left) or user-defined (right) floor plans.}
        \label{fig:search_result}
    \end{figure}

\noindent \textbf{Applications.} Our method enabled searching for similar and dissimilar floor plans of a given floor plan in interactive speeds. With our method, exhaustively comparing a floor plan to all others in the dataset took about 8.9 seconds, while using LayoutGMN to compute all the similarity scores took about 296.8 seconds. Users can also insert new floor plans into the dataset. To calculate the coordinate of the new data point, we use LayoutGMN to calculate its similarity scores to all existing floor plans in the dataset, and solve the optimization (Equation~\ref{equ:optimization}) again but with all existing data points' coordinates fixed. Figure \ref{fig:search_result} shows multiple results with existing floor plans in the dataset and novel designs by users as inputs. In general, we found that it is easier to find designs similar to existing entries in the dataset rather than novel designs specified by the user. We think the reason is due to the limited diversity of the dataset. Finally, our method enabled us to detect and prune nearly redundant entries in the original RPLAN dataset. We found 4419 redundant floor plans that are only 50 pixels or less different to others (layout images are of a 256x256 resolution).

\noindent \textbf{User Study.} We conducted a questionnaire-based survey of 25 correspondents. Our questionnaire is divided into two parts: ranking and judgment. By calculating the rank correlation, our system obtained a positive correlation result of 63.6\% in the ranking part. In the judgment section, there is an 85\% pass rate.

\section{Conclusion}
    
We propose a method to encode the similarity measurements of LayoutGMN in a common low-dimensional space to facilitate much more efficient exploration. For future work, we would like to explore different ways to encode the similarity scores such as weighted sums of Euclidean distances and cosine similarity.


\bibliographystyle{eg-alpha-doi} 
\bibliography{egbibsample}       

\end{document}